\documentstyle[aps,epsfig,floats,preprint]{revtex}
\begin{document}
\draft
\preprint{\hbox{SOGANG-HEP 272/00}}
\title{Brane-World Black Holes in Randall-Sundrum Models}
\author{Won Tae Kim\footnote{electronic address:wtkim@ccs.sogang.ac.kr},
  John J. Oh\footnote{electronic address:john5@string.sogang.ac.kr},
  Marie K. Oh\footnote{electronic address:kmoh@string.sogang.ac.kr},
  and Myung Seok Yoon\footnote{electronic address:younms@physics.sogang.ac.kr}}
\address{Department of Physics and Basic Science Research Institute,\\
         Sogang University, C.P.O. Box 1142, Seoul 100-611, Korea}
\date{\today}
\maketitle
\begin{abstract}
We study brane-world black holes from Randall-Sundrum(RS) models in
($D+1$)-dimensional anti-de Sitter spacetimes. The solutions are
directly obtained by using a slightly generalized RS metric ansatz in
$D+1$ dimensions. The metric of the brane world can be described by the
Schwarzschild solution promoted to the black cigar solution in $D+1$
dimensions, which is compatible with the recently suggested black
cigar solution in $D=4$. Furthermore, we show that the Ricci flat
condition for the brane can be easily derived from the effective
gravity defined on the brane by using the RS dimensional
reduction. Especially, it is shown that in two dimensions the
effective gravity on the brane is described by the Polyakov action.
\end{abstract}

\newpage
\section{Introduction \hfill{}}\label{intro}

Recently, there has been much interests in the idea that our universe
may be embedded in some higher-dimensional spacetimes. It has been
suggested that the gauge hierarchy problem can be resolved in higher
dimensions~\cite{add,aadd}. This framework unifies gravitation
and gauge interactions at the weak scale and explains their
hierarchy in our spacetime. However, this leads to
another hierarchy between weak scale and compactification one.

On the other hand, in order to solve the hierarchy problem Randall and
Sundrum(RS) have proposed a two-brane model called RS1 model
involving a small and
curved extra dimension which is a slice of anti-de Sitter (AdS)
spacetime~\cite{rs1}. The negative tension brane is regarded 
as our universe and the hierarchy between physical scales naturally
appears in our brane. Furthermore, RS have studied a single-brane
model(RS2 model) by taking $r_{c} \rightarrow \infty$, where $r_{c}$ is a
radius of the extra dimension~\cite{rs2}. In these RS models, the
nonfactorizable metric with $-\pi r_c \le y \le \pi r_c$  
\begin{equation}
  \label{rs1}
  ds^2 = e^{-2k |y|} \eta_{\mu\nu} dx^{\mu} dx^{\nu} +  dy^2, 
\end{equation}
is essential as a static vacuum, which is different from that of the
conventional Kaulza-Klein(KK) style in that the extra coordinate
is associated with the conformal factor. On the other hand, the
cosmological aspects of the RS model are intensively studied in
Ref. \cite{cos}. The linearized gravity
on the RS brane world is also treated in Refs. \cite{rs2,lr,gkr} and
the dilatonic brane-world model 
are covered in Ref. \cite{dil}. 

Apart from the RS model, the static AdS domain wall has been found as
BPS domain walls of $D=4$, $N=1$ supergravity theoris in
Ref. \cite{cgr}, which is relevant for RS model in a special case. A
further study on the global spacetime and non-static solution in this
subject has been shown in Ref. \cite{cve}, and the $D$-dimensional
approach has been studied in Ref. \cite{cw}. 

As for black holes, Chamblin, Hawking, and Reall(CHR) \cite{chr} have
proposed that a nonrotating and uncharged black hole on the domain wall is
described by a black cigar solution in five dimensions. 
The curvature singularity and black hole mass have been
discussed in detail. In Ref. \cite{chr}, 
the vacuum metric (\ref{rs1}) is generalized by substituting
$\eta_{\mu \nu}$ with $g_{\mu \nu}$ under the Ricci flat condition,
and find the brane-world solution. Note that black hole
solutions in intersecting domain wall backgrounds has been studied in
Ref. \cite{cce}. 

In the four-dimensional RS model, brane-world black holes
and their thermodynamics \cite{ehm}, and a comparison with BTZ black holes and
black strings on brane-world black holes \cite{ehm2} have been studied. 
Furthermore, in Ref. \cite{ehm3}, aspects of evaporation in brane-world black holes
have been discussed.

In Sec. \ref{bwbh}, we would like to obtain the brane-world black hole
solutions from ($D+1$)-dimensional AdS gravity with $(D-1)$-branes
and show that they satisfy the Ricci flat condition in all dimensions. In
Sec. \ref{etdr}, we find the effective gravity of the brane in terms of the RS
dimensional reduction. In our calculations, 
the Ricci flat condition naturally appears as an equation of motion even though
there exists some scalar field couplings in the effective
action, so that the desirable brane-world solution can be easily obtained.
On the other hand, the single brane limit is well defined,
and the effective gravity just becomes Einstein
gravity. Furthermore, for the case of three dimensions two-dimensional
effective gravity is described by the well-known Polyakov
action. In Sec. \ref{diss}, we discuss our results comparing with
the RS2 model.

\section{Brane-World Black Holes in (D+1)-Dimensions \hfill{}} \label{bwbh}

Let us now consider the first RS (RS1) model in $(D+1)$ dimensions, 
\begin{eqnarray}
S_{(D+1)} &=& \frac{1}{2 \kappa^2_{(D+1)}} \int d^D x \int dy
  \sqrt{-g_{(D+1)}} \left[ R^{(D+1)} + D(D-1)k^2 \right] \nonumber \\
& & - \int d^D x \sqrt{-\tilde{g}_{(+)}} \lambda_{(+)} -  \int d^D x
  \sqrt{-\tilde{g}_{(-)}} \lambda_{(-)}, \label{2baction}
\end{eqnarray}
where $\lambda_{(+)}$ and $\lambda_{(-)}$ are tensions of the branes
at $y=0$ and at $y=\pi r_c$, respectively, $D(D-1)k^2$ is a
cosmological constant, and $\kappa_{(D+1)}^2=8\pi G_{N}^{(D+1)}$.
Note that we use $M,N,K, \cdots = 0, 1, \cdots, D$
for $(D+1)$-dimensional spacetime indices and $\mu, \nu, \kappa,
\cdots = 0, 1, \cdots, D-1$ for branes. We assume  
orbifold $S^1/Z_{2}$ which has a periodicity in the
extra coordinate $y$ and identify $-y$ with $y$. Two singular points
on the orbifold are located at $y=0$ and $y=\pi r_c$, and two
($D-1$)-branes are placed at these points, respectively. Note that
$\tilde{g}_{\mu\nu}^{(+)}$ and $\tilde{g}_{\mu\nu}^{(-)}$ are defined as 
\begin{eqnarray}
  \tilde{g}_{\mu \nu}^{(+)} &\equiv& g_{\mu\nu}^{(D+1)}(x^{\mu},y=0), 
    \nonumber \\ 
 \tilde{g}_{\mu \nu}^{(-)} &\equiv& g_{\mu\nu}^{(D+1)}(x^{\mu}, y=\pi
    r_c).  \label{metrdef}
\end{eqnarray}
We assume a generalized metric as
\begin{equation}\label{metrican}
ds^2_{(D+1)} = e^{\sigma(y) \Phi(x)} g_{\mu\nu}^{(D)}(x) dx^{\mu}dx^{\nu}
+ T^2 (x) dy^2, 
\end{equation}
where the moduli field $T(x)$ is different from $\Phi (x)$
for the present. From Eq.~(\ref{2baction}), the equations of motion
are given as 
\begin{equation}
  \label{einseqnss}
  G_{MN}^{(D+1)} = T_{MN}^{(D+1)}.
\end{equation}
By using the metric (\ref{metrican}), 
the Einstein tensors are calculated as
\begin{eqnarray}
  G_{\mu\nu}^{(D+1)} &=& G_{\mu\nu}^{(D)} - \frac{1}{T}
    \left(\nabla_{\mu} \nabla_{\nu}  
    T - g_{\mu\nu}^{(D)} \Box T \right) - \frac{1}{2} \sigma (D-2)
    \left(\nabla_{\mu}\nabla_{\nu} \Phi - g_{\mu\nu}^{(D)} \Box \Phi \right)
    \nonumber \\ 
  & & + \frac{1}{2T} \sigma \left(\nabla_{\mu}T \nabla_{\nu}\Phi +
    \nabla_{\mu}\Phi \nabla_{\nu}T + (D-3) g_{\mu\nu}^{(D)}
    \nabla_{\rho}T\nabla^{\rho} \Phi \right) \nonumber \\ 
  & & + \frac{1}{8} \sigma^2 (D-2) \left( 2 \nabla_{\mu}\Phi
    \nabla_{\nu}\Phi + (D-3)g_{\mu\nu}^{(D)} (\nabla \Phi)^2 \right)
    \nonumber \\ 
  & & + \frac{\Phi}{8T^2}(D-1) e^{\sigma \Phi} g_{\mu\nu}^{(D)} \left(
    \Phi D(\partial_{y} \sigma)^2 + 4\partial_{y}^2 \sigma \right),
    \label{einseq1} \\ 
  G_{\mu y}^{(D+1)} &=& - \frac{1}{2} (D-1) \left( \partial_{\mu} \Phi -
    \frac{\Phi}{T} \partial_{\mu}T\right) \partial_{y} \sigma,
    \label{einseq2} \\
  G_{yy}^{(D+1)} &=& - \frac{1}{2} T^2 e^{-\sigma \Phi} \left( R^{(D)}
    - \sigma(D-1) \Box \Phi - \frac{1}{4} \sigma^2 (D-1)(D-2) (\nabla
    \Phi)^2\right. \nonumber \\ 
  & & - \left. \frac{\Phi^2}{4T^2} D(D-1) e^{\sigma \Phi} (\partial_{y}
    \sigma)^2 \right), \label{einseq3}
\end{eqnarray}
and the stress-energy tensor is explicitly written as
\begin{eqnarray}
  \label{stress}
  T_{MN}^{(D+1)} = \frac{1}{2} g_{MN}^{(D+1)} D(D-1)k^2 &+& \kappa_{(D+1)}^2 \frac{\sqrt{-\tilde{g}_{(+)}}}{\sqrt{-g_{(D+1)}}}
  \lambda_{(+)} \delta(y) \tilde{g}_{\mu\nu}^{(+)}
  \delta_{M}^{\mu} \delta_{N}^{\nu}\nonumber \\ &+&
  \kappa_{(D+1)}^2 \frac{\sqrt{-\tilde{g}_{(-)}}}{\sqrt{-g_{(D+1)}}}
  \lambda_{(-)} \delta(y-\pi r_c) \tilde{g}_{\mu\nu}^{(-)}
  \delta_{M}^{\mu} \delta_{N}^{\nu} 
  .
\end{eqnarray}
Since the ($\mu y$)-component of Eq. (\ref{stress}) vanishes, from Eq. (\ref{einseqnss}) we obtain
the following relation, 
\begin{equation}
  \label{consteq}
  G_{\mu y}^{(D+1)} = - \frac{1}{2} (D-1) \left( \partial_{\mu} \Phi -
  \frac{\Phi}{T} \partial_{\mu}T\right) \partial_{y} \sigma = 0.
\end{equation}
It is interesting to note that the case of $\sigma(y)={\rm constant}$
corresponding to a factorizable geometry describes a conventional
Kaluza-Klein(KK) theory without a KK gauge field $A_{\mu}(x)$. For our
nonfactorizable metric (\ref{metrican}), we take $\sigma(y) = -2k|y|$ by considering RS vacuum metric (\ref{rs1}), then Eq.~(\ref{consteq}) yields
\begin{equation}
  \label{constraint}
  \Phi(x) = T(x).
\end{equation}

\begin{figure}[htbp]
  \begin{center}
    \leavevmode
    \centerline{
        \epsfig{figure=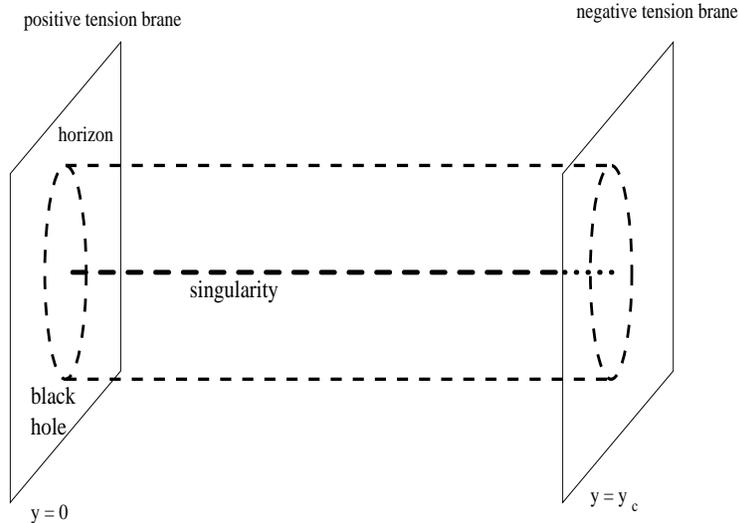, width=10cm, height=7cm}
        }
    \caption{The black hole can be formed by collapsing matters on the
       brane and its singularities exist along the extra dimension.}
    \label{fig:double}
  \end{center}
\end{figure}

In the ($\mu\nu$)-components of Eq. (\ref{einseqnss}), there exist
discontinuities resulting from the delta functional source due to the
presence of brane tensions at $y=0$ and at $y=\pi r_c$. At this
stage, we consider junction conditions and integrate out the Einstein
equation near the branes, 
\begin{eqnarray}
& &\int_{0-\epsilon}^{0+\epsilon}dy G_{\mu\nu}^{(D+1)} =
  \int_{0-\epsilon}^{0+\epsilon} dy T_{\mu\nu}^{(D+1)}, \nonumber \\ 
& &\int_{\pi r_c-\epsilon}^{\pi r_c+\epsilon}dy G_{\mu\nu}^{(D+1)} =
  \int_{\pi r_c-\epsilon}^{\pi r_c+\epsilon}dy T_{\mu\nu}^{(D+1)}. 
  \label{integrated} 
\end{eqnarray}
The jump along the extra coordinate near the ($D-1$)-branes gives a relation,
\begin{equation}\label{rel2b}
  \lambda_{(+)} = - \lambda_{(-)} = \frac{2(D-1)}{\kappa_{(D+1)}^2} k,
\end{equation}
where we note that the branes at $y=0$ and at $y=\pi r_c$ have a positive
tension ($\lambda_{(+)}$) and a negative one ($\lambda_{(-)}$),
respectively. Using the relation
(\ref{rel2b}) for this brane model, the equation of motion
(\ref{einseqnss}) is explicitly given as  
\begin{eqnarray}
& & R_{\mu \nu}^{(D)} - \frac12 g_{\mu\nu}^{(D)} R^{(D)} - \frac{1}{T}
  \left[ \nabla_{\mu} \nabla_{\nu} T - g_{\mu\nu}^{(D)} 
  \Box T\right] + \frac12 k^2 y^2 (D-2) \left[
  2\nabla_{\mu}T\nabla_{\nu}T + (D-3) g_{\mu\nu}^{(D)} (\nabla T)^2
  \right] \nonumber \\
& & \qquad+ k|y|(D-2)\left[ \nabla_{\mu} \nabla_{\nu} T -
  g_{\mu\nu}^{(D)} \Box T\right] - \frac{k|y|}{T}\left[2
  \nabla_{\mu}T\nabla_{\nu}T + (D-3) g_{\mu\nu}^{(D)} (\nabla T)^2
  \right]=0, \label{eqnmot1} \\  
& & R^{(D)} + 2k|y|(D-1) \Box T - k^2 y^2 (D-1)(D-2) (\nabla T)^2 =
  0. \label{eqnmot2} 
\end{eqnarray}
Trace of Eq. (\ref{eqnmot1}) and Eq. (\ref{eqnmot2}) 
give the following simple equations, 
\begin{eqnarray}
  \label{reduceqn}
  R^{(D)} + k|y|(D-1) \Box T = 0, \nonumber \\
  \Box T - k|y|(D-2) (\nabla T)^2 = 0.
\end{eqnarray}
In Eq.~(\ref{reduceqn}), as a simple constant solution of $T(x)$, we set $T(x)=1$, then metric solution $g_{\mu\nu}^{(D)}$ should be determined by a condition, 
\begin{equation}
\label{ricciflat}
R^{(D)} = 0.
\end{equation}
Combining $R^{(D)} = 0$ and $T(x)=1$ with Eq. (\ref{eqnmot1}), the Ricci flat condition, $R_{\mu\nu}^{(D)} = 0$ introduced in Ref. \cite{chr},
is obtained from the equations of motion. 
>From this condition (\ref{ricciflat}), 
it is natural to consider the $D$-dimensional
Schwarzschild black hole solution for uncharged nonrotating case
as a brane solution, 
\begin{equation}
  \label{d4ss}
ds^2 = e^{-2k|y|}\left[ - \left(1-\frac{2M}{r}\right)
dt^2 + \left(1-\frac{2M}{r}\right)^{-1} dr^2 + r^2 d\Omega^2_{(D-2)}
\right] + dy^2, 
\end{equation}
where $d\Omega^2_{(D-2)}$ is a metric of unit $(D-2)$-sphere. 
For $D=4$, it is a ``black cigar'' solution 
which describes a black hole placed on the hypersurface 
at the fixed extra coordinate $y$ \cite{chr}. 
The black hole horizon is $r_{H} = 2M$ \cite{chr}, which is shown in FIG. \ref{fig:double}. 

On the other hand, the Arnowitt-Deser-Meisner(ADM) mass 
$\tilde{M}$ of 
the brane-world black hole measured on the brane is given as
\begin{equation}
  \label{adm}
  \tilde{M} = M e^{-ky_{0}},
\end{equation}
where $y_{0}$ is $0$ and $\pi r_{c}$ which are $y$-coordinates of the
positive and the negative tension branes, respectively. 

The Ricci scalar and the square of the
Ricci tensor are constant, while the square of Riemann tensor 
is written as
\begin{equation}
  \label{rieman}
  R_{MNKL}R^{MNKL} = 10 k^4 + \frac{12 M^2 e^{4ky}}{r^6},
\end{equation}
especially for $D=4$. It can be shown that in Eq.~(\ref{rieman}),
there exists a curvature singularity at $r=0$ and this is extended 
along the $y$-coordinate such as a string in FIG. \ref{fig:double}. The range of $y$ spans from
$0$ to $y_c = \pi r_c$ for the RS1 model, while for the RS2 model,
$y_{c}$ goes to infinity. Therefore, for $y_{c} \rightarrow \infty$,
the square of Riemann tensor diverges for finite $r$ and the ADM mass
would be exponentially suppressed as $\tilde{M} = M e^{-ky_{c}}
|_{y_{c}\rightarrow\infty} \rightarrow 0$ on the negative tension
brane for the RS2 model. 

\section{Effective Theory from the Dimensional Reduction \hfill{}} \label{etdr}

We now study $D$-dimensional effective theory on the brane by using
the dimensional reduction with the generalized metric (\ref{metrican}) of
RS models and by including the constraint (\ref{constraint}) to be
consistent with ($D+1$)-dimensional Einstein equation and find an
explicit action describing the brane world. At first
sight, a candidate for the theory seems to be Einstein gravity
because of the Ricci flat condition. However, the resulting action
looks like a dilaton gravity satisfying the Ricci flat condition. 
Using the metric (\ref{metrican}), a Ricci scalar can be expressed as 
\begin{eqnarray}
  \label{ricci}
R^{(D+1)} &=& e^{-\sigma \Phi} \left[ R^{(D)} - \sigma (D-1) \Box
    \Phi - \frac{2}{T} \Box T - \frac{\sigma}{T}(D-2)\nabla_{\mu}\Phi
    \nabla^{\mu} T \right.\nonumber \\ 
& & - \left. \frac{1}{4} \sigma^2 (D-1)(D-2) (\nabla \Phi)^2 -
    \frac{\Phi^2}{4 T^2} D(D+1) e^{\sigma \Phi} (\partial_{y}
    \sigma)^2 - \frac{\Phi}{T^2} D e^{\sigma \Phi} 
    \partial_{y}^2 \sigma \right]. 
\end{eqnarray}
Inserting Eq.~(\ref{ricci}) into Eq.~(\ref{2baction}) for
$\sigma(y)=-2k|y|$ and the constraint (\ref{constraint}), we obtain
\begin{eqnarray}
S_{(D+1)} &=& \frac{1}{2{\kappa_{(D+1)}}^2} \int d^{D}x
  \sqrt{-g_{(D)}}T \int_{-\pi r_c}^{\pi r_c} dy  e^{-(D-2)k|y| T} \left[
  R^{(D)} + 2k|y| (D-1) \Box T \right.\nonumber \\ 
& & - \frac{2}{T}
  \Box T + \frac{2k|y|}{T}(D-2)(\nabla T)^2 - k^2 y^2 (D-1)(D-2)
  (\nabla T)^2 \nonumber\\ 
& & - \left. D(D+1) e^{-2k|y|T}k^2 +
  \frac{4}{T}k D e^{-2k|y|T} (\delta(y)-\delta(y-\pi r_c)) + e^{-2k|y| T}
  D(D-1)k^2 \right]\nonumber \\ 
& & - \int d^{D}x \sqrt{-g_{(D)}}
  \lambda_{(+)} - \int d^{D}x \sqrt{-g_{(D)}}e^{-D k\pi r_c T}
  \lambda_{(-)}. \label{genactionred}
\end{eqnarray}
After integrating out along the extra dimension $y$ from $-\pi r_c$ to
$\pi r_c$, the $D$-dimensional effective action is now obtained as
\begin{eqnarray}
  \label{2beffecta}
  S_{(D)} &=& \frac{1}{2 \kappa_{(D+1)}^2 k} \int d^{D}x \sqrt{-g_{(D)}}
  \left[ \frac{2}{(D-2)} \left( 1- e^{-(D-2) \phi} \right) R^{(D)} -
  2(D-1) e^{-(D-2)\phi} (\nabla \phi)^2 \right. \nonumber \\
   &+& \left.\left( 4(D-1) k^2 - 2 \kappa^2_{(D+1)} k \lambda_{(+)}\right)- e^{-D\phi} \left(4(D-1) k^2 + 2 \kappa^2_{(D+1)} k \lambda_{(-)}\right)\right],
\end{eqnarray}
where we redefined $\phi(x) \equiv \pi k r_c T(x)$. If we use junction condition (\ref{rel2b}) in Eq. (\ref{2beffecta}), then the resulting effective action can be obtained as
\begin{equation}
  \label{2beffectac}
  S_{(D)} = \frac{1}{2 \kappa_{(D+1)}^2 k} \int d^{D}x \sqrt{-g_{(D)}}
  \left[ \frac{2}{(D-2)} \left( 1- e^{-(D-2) \phi} \right) R^{(D)} -
  2(D-1) e^{-(D-2)\phi} (\nabla \phi)^2\right] , 
\end{equation}
The action (\ref{2beffectac}) is reminiscent of the low-energy
string theory with some modified dilaton coupling, however,
it is different in that the Ricci flat condition is satisfied
through the equations of motion.  

>From the effective action (\ref{2beffectac}), 
equations of motion are
derived as 
\begin{eqnarray}
  \label{effeom}
& & \left[ \left(1-e^{-(D-2)\phi}\right)
  (R_{\mu\nu}^{(D)} - \frac{1}{2} g_{\mu\nu}^{(D)} R_{(D)}) + (D-2)^2
  e^{-(D-2)\phi} \nabla_{\mu}\phi\nabla_{\nu}\phi\right.\nonumber \\ 
& &\qquad - \left.(D-2) e^{-(D-2)\phi} \nabla_{\mu}\nabla_{\nu} \phi 
  - g_{\mu\nu}^{(D)} (D-2)^2 e^{-(D-2)\phi} (\nabla\phi)^2 +
  g_{\mu\nu}^{(D)} (D-2) e^{-(D-2)\phi} \Box \phi \right]\nonumber \\
& & \qquad + (D-1)(D-2) e^{-(D-2)\phi}
  \left(\frac{1}{2}g_{\mu\nu}^{(D)}(\nabla \phi)^2 - \nabla_{\mu}\phi
  \nabla_{\nu}\phi\right)=0, \label{effeom1}\\ 
& &R^{(D)} - (D-1)(D-2)(\nabla \phi)^2 + 2(D-1) \Box \phi =
  0,\label{effeom2} 
\end{eqnarray}
and the trace of Eq. (\ref{effeom1}) gives 
\begin{equation}
  \label{traceeffeom1}
   R^{(D)} - e^{-(D-2)\phi}\left( R^{(D)} - (D-1)(D-2)(\nabla \phi)^2
   + 2(D-1) \Box \phi\right)=0. 
\end{equation}
Combining Eqs.~(\ref{effeom2}) and (\ref{traceeffeom1}), the condition $R^{(D)} = 0$ is obtained, which is consistent with
the previous Eq.~(\ref{ricciflat}) from the ($D+1$)-dimensional
analysis. Here, if we assume a constant background $\phi$ solution and use
$R^{(D)}=0$, then the Ricci flat condition shown in
the previous section can be reproduced in Eq. (\ref{effeom1}).

As for the special case of $D=2$, we take the limit of $D \rightarrow 2$
in the action (\ref{2beffectac}), then two-dimensional brane world is
governed by
\begin{equation}
  \label{d=2limit}
  S_{(D=2)} = \frac{1}{\kappa_{(3)}^2 k} \int d^2x \sqrt{-g_{(2)}}
  \left[ \phi R^{(2)} - (\nabla \phi)^2 \right].
\end{equation}
Integrating out $\phi(x)$, the well-known two-dimensional Polyakov
action \cite{pol} is obtained, 
\begin{equation}
 \label{polyakov}
S_{2D} = -\frac{1}{4 \kappa_{(3)}^2 k} \int dx^2 \sqrt{-g_{(2)}}
 R^{(2)}\frac{1}{\Box} R^{(2)}.
\end{equation}
This result is of interest in that the Polyakov action is derived from
the classical three-dimensional brane-world model. From the beginning,
if one fixes $D=2$, then the same result comes out.

\section{Discussions \hfill{}}\label{diss}

Now, it seems to be appropriate to comment on a single
brane model of RS2 model \cite{rs2}, 
the similar analysis to the RS1 model
can be applied to the RS2 model, however, 
for simplicity, 
we take the
limit of $r_{c} \rightarrow \infty$ corresponding to $\phi(x)
\rightarrow \infty$ in Eq.~(\ref{2beffectac}). Then, the resulting
action gives a desirable result as 
\begin{equation}
  \label{reducedact}
  S_{(D)} = 
\frac{1}{(D-2)\kappa_{(D+1)}^2 k} \int d^{D}x \sqrt{-g_{(D)}} R^{(D)}.
\end{equation}
It is the $D$-dimensional Einstein-Hilbert action which yields a
$D$-dimensional Schwarzschild black hole solution. For the present RS2
model, if $y \rightarrow \infty$, the ADM mass (\ref{adm}) is
exponentially suppressed and the black hole horizon shrinks to zero at
the AdS horizon.

Finally, the effective mass scale $M_{D}^2$ in $D$ dimensions from Eq. (\ref{2beffectac})
can be defined as  
\begin{equation}
  \label{coupling}
  M_{(D)}^2 = \frac{2 M_{(D+1)}^3}{(D-2)k} 
\left( 1 - e^{-\frac{(D-2)}{2}\phi} \right),
\end{equation}
where $M_{(D+1)}^3$ is a mass scale in $(D+1)$-dimensional spacetimes which is defined as $1/{\kappa_{(D+1)}^2}$. Note that the relation (\ref{coupling}) is consistent with the result of the original cosmological
derivation \cite{rs1} for the constant background $\phi=\pi k r_c$ in
$D=4$. 

In summary, we have studied the brane-world black hole solutions in the 
RS model. The RS dimensional reduction with the junction conditions 
gives the interesting brane-world gravity. 
The Ricci flat condition appears as one of the reduced equations of motion
and the solutions are easily solved.
As a comment, the effective gravity
might reflect more or less holographic properties in that
especially for $D=2$ the Polyakov induced gravity is
a conformal field theory with the central charge $c=\frac{24\pi}{\kappa_{(3)}^2 k}$ in the spirit of AdS/CFT correspondence \cite{mal}, so
that it would be interesting to study this issue in more detail.

\vspace{1cm}

{\bf Acknowledgments}\\
We would like to thank P. P. Jung for helpful discussions. We are
grateful to M. Cveti${\check{c}}$ for useful comments on domain walls,
and would like to S. D. Odintsov, J. Erlich, and R. C. Myers for
helpful remarks on our work. This work was supported by the Ministry of Education, Brain Korea 21
Project No. D-0055, 1999.


\end{document}